\PassOptionsToPackage{svgnames}{xcolor}
\documentclass[12pt,reqno]{article}
\usepackage{base}
\usepackage[utf8]{inputenc}
\usepackage{theorems}
\usepackage[numbers,sort&compress]{natbib}
\usepackage{amsthm, amsmath, amsfonts, amssymb, amscd, mathtools, youngtab, euscript, mathrsfs, verbatim, enumerate, multicol, multirow, bbding, color, babel, esint, geometry, tikz, tikz-cd, tikz-3dplot, array, enumitem, hyperref, thm-restate, thmtools, datetime, graphicx, tensor, braket, slashed, standalone, pgfplots, ytableau, subfigure, wrapfig, dsfont, setspace, wasysym, pifont, float, rotating, adjustbox, pict2e,array}
\usepackage{amsfonts}
\usepackage{longtable}
\usepackage{tabularx}

\usepackage{caption}

\usepackage{booktabs}
\usepackage{amsmath}
\usepackage{etoolbox}
\usepackage{makecell}
\usepackage{physics}
\usepackage{blindtext}
\usepackage{mathalpha}
\usepackage[all]{xy}

\usepackage[utf8]{inputenc}
\usepackage[normalem]{ulem}
\makeatletter
\numberwithin{figure}{section}
\usetikzlibrary{arrows, positioning, decorations.pathmorphing, decorations.pathreplacing, decorations.markings, matrix, patterns}
\hypersetup{colorlinks=true}
\hypersetup{linkcolor=black}
\hypersetup{citecolor=black}
\hypersetup{urlcolor=black}
\usepackage{physics}
\usepackage{amsmath}
\usepackage{tikz}
\usepackage{mathdots}
\usepackage{yhmath}
\usepackage{cancel}
\usepackage{color}
\usepackage{siunitx}
\usepackage{array}
\usepackage{multirow}
\usepackage{amssymb}
\usepackage{gensymb}
\usepackage{tabularx}
\usepackage{extarrows}
\usepackage{booktabs}
\usetikzlibrary{fadings}
\usetikzlibrary{patterns}
\usetikzlibrary{shadows.blur}
\usetikzlibrary{shapes}
\usepackage{framed}

\usepackage{xcolor}
\definecolor{blue}{RGB}{33, 118, 199}

\definecolor{green}{RGB}{0, 128, 0}

\definecolor{red}{RGB}{230, 0, 20}

\usepackage{hyperref}
\hypersetup{
colorlinks=true,
urlcolor= blue,
citecolor=blue
}

\definecolor{darkgreen}{RGB}{50,150,0}

\newcommand{\R}{\mathbb{R}}
\newcommand{\Z}{\mathbb{Z}}
    {\endtcolorbox}
\usepackage{mdframed}
\mdfdefinestyle{statementstyle}{
   frametitlerule = true,
   frametitlefont = {\normalfont\bfseries},
   frametitlebackgroundcolor = blue!10,
}
\mdtheorem[style=statementstyle]{statement}{Type I' supergravity}

\begin{document}
\begin{titlepage}
\hfill \\
\vspace*{15mm}
\begin{center}

{ \Large \textbf{Non-BPS path to the string lamppost}}

\vspace*{15mm}
\vspace{2cm}
{\large
Alek Bedroya{$^1$}, Sanjay Raman{$^2$}, Houri-Christina Tarazi{$^1$}\\}
\vspace{.6cm}
{ $^1$ Department of Physics, Harvard University, Cambridge, Massachusetts, USA}\par
\vspace{.2cm}
{ $^2$ Department of Physics, MIT, Cambridge, Massachusetts, USA}\par\vspace{-.3cm}

\vspace{1cm}
\begin{abstract}
We provide further motivation for the string lamppost principle in 9d supergravities. Using a blend of ideas which includes Swampland conjectures, finiteness of black hole entropy, and classification of SCFTs, we show that infinite distance limits that keep BPS states heavy must decompactify to type IIA supergravity on an interval. Without relying on string theory, we provide bottom-up explanations for various UV features of the theory, such as the physics near the orientifold branes and the worldvolume theories of different stacks of non-perturbative 8-branes. We also provide a Swampland argument for the countability of the number of inequivalent string limits up to dualities which is a strong result with applications beyond this work.
\end{abstract}

\end{center}
\end{titlepage}
\tableofcontents

\section{Introduction}

Low-energy field theories provide a useful window into studying quantum field theories, with the hope that one can always study higher energies by increasing the cut-off and integrating-in the massive states. However, in quantum gravity, this logic always breaks down due to the existence of black holes. Gravity screens the highly massive states of the theory and turns them into black holes\footnote{For a review see \cite{Bedroya:2022twb} and references therein.}. This suggests that the massive states of the theory play an important role in defining a gravitational theory. 

One can hope to access and study the massive states of the theory by changing the parameters which control their masses to study them in a regime where they become light and weakly coupled. However, any equation governing those states in the weakly coupled regime generally receives large corrections in the strongly coupled regime. BPS states avoid this problem because their masses are protected by supersymmetry; they have therefore  played an important role in understanding black holes in string theory \cite{Strominger:1996sh,Maldacena:1997de}. However, in this paper, we take a different approach and focus on the regime of moduli space where all light massive particles are non-BPS. We show that, using some Swampland conjectures, even the structure of such less controlled limits can be highly constrained. The Swampland conjectures are conjectural consistency conditions that aim to delineate the landscape of low-energy theories in quantum gravity from those that do not have a quantum gravity UV completion (i.e. Swampland) \cite{Vafa:2005ui}. Over the last decade, much work has been done to determine and understand these conditions (see \cite{Agmon:2022thq} for a recent review). 

 In addition to being supported by string theory, the Swampland conjectures have varying levels of support from fundamental principles (unitarity, holography, etc.) and consistency amongst themselves. An important question is whether the support from string theory could be misleading because there could be other theories of quantum gravity. A good way to approach this concern is to  reduce string theory to more basic principles to be examined more carefully. In other words, we would like to understand what is the minimal set of assumptions that will reproduce the same results as string theory? The postulate that string theory is the unique theory of quantum gravity is known as the String Lamppost Principle (SLP). In this work, we take a new step toward showing that this principle is closely related to simpler Swampland conditions.

 In recent years a lot of work has been done to use Swampland conjectures and show the uniqueness of string theory features in supergravities. In particular, it was shown in \cite{Kim_4DSYM} that in theories with 16 supercharges, the rank of the gauge group is bounded by $r_G\leq 26-d$, which is satisfied and can be saturated in string theory. Consequent works \cite{Montero_SLP} extended that result and used Swampland conjectures to prove a refined statement that in 9 and 8 dimensions, the possible gauge group ranks satisfy $r_G=1\mod 8 $ and $r_G=2\mod 8$, respectively, which all have some string theory realization \cite{Dabholkar_Orientifolds, Aharony_2007}. In subsequent works \cite{Cveti_2020,Hamada_recon}, the bottom-up argument for such features of string theory were taken a step further  in \cite{Bedroya21}, the authors used Swampland conjectures to derive a list of gauge algebras in $d>6$ supergravities\footnote{The $d=7$ case assumed a classification of 3d SCFTs which lacks a bottom-up argument outside string theory.} and showed that they identically match the algebras realized in string theory. 

Similar efforts to constraint the low energy physics in theories with fewer supercharges was conducted in
\cite{Tarazi:2021duw,Kim:2019vuc, Katz:2020ewz, Martucci:2022krl}.

Despite the remarkable progress in motivating the SLP by finding the gauge enhancements from Swampland conditions, there are other features of string theory that lack a bottom-up explanation. For example, string theory gives us a detailed account of the light weakly coupled states in any corner of the moduli space. In this note, we aim to expand on the existing work to demonstrate this aspect of the SLP for 9d supergravities with minimal supersymmetry.

We use a strong version of the sharpened distance conjecture \cite{Etheredge:2022opl}, which claims that as we go to any infinite distance limit in the moduli space, there is a tower of weakly coupled light particles with masses scaling as $m^2 \sim e^{-\lambda \Vert \phi \Vert}$, where $\Vert \cdot \Vert$ is the canonical distance travelled in the moduli space. Moreover, if the coefficient in the exponential satisfies $\lambda = 1/\sqrt{d-2}$ the light particles must be states of a light string. Otherwise, $\lambda > 1/\sqrt{d-2}$, and the tower is a KK tower. This conjecture is a stronger version of distance conjecture \cite{Etheredge:2022opl} and follows from the emergent string conjecture \cite{Lee:2019wij} (see \cite{Agmon:2022thq} for derivation). Using this conjecture, we show that the number of string limits must be countable. This statement implies that limits in the moduli space of 9d theories with no light BPS particles must decompactify to IIA supergravity on an interval with certain non-perturbative branes on the interval. Moreover, we find the gauge symmetry living on those non-perturbative 8-branes, which provides a bottom-up derivation of the worldvolume theory of D8 branes. Finally, our work fixes the charge lattice for all 9d theories, regardless of the rank of the gauge group. We also briefly discuss extensions of our approach to fewer dimensions.

\section{Infinite distance limits in $d > 6$}

In a gravitational theory, large field excursions towards the boundary of the moduli space are expected to reveal some new features. In this work we are going to focus at such limits that can be reached only at infinite distance in the moduli space\footnote{In lower supersymmetries or in conformal fixed points new massless modes can correct the metric of the moduli space and bring a formally infinite distance limit to a finite distance.}. 
According to the Distance Conjecture, these limits are proposed to have an emergent tower of massless states \cite{Ooguri_DC}, signaling the breaking of the current description and the emergence of a dual weakly-coupled effective description. The simplest example of such a tower is the KK tower associated with decompactification. For example, M-theory on a circle has the $R\to  0$ limit where light towers of membranes wrapping the circle appear (IIA strings) and the $R\to \infty$ where the light tower consists of the KK states signaling the decompactification. The emergent string conjecture, a refinement of the Distance Conjecture, says that at these infinite distance limits the theory, there is either a decompactification or an asymptotically tensionless string \cite{Lee:2019wij}. 

In this work we are interested in theories with 16 supercharges in 9 dimensions and  in infinite distance limits where no BPS particle becomes light. 

Let us first look at  string theory and identify such infinite distance limits. In 10 dimensions with 16 supercharges, there is only one scalar (the dilaton) and there are no BPS particles. The first nontrivial place to begin our study is therefore with 9 dimensional theories. Consider in particular theories with maximal rank (rank 17) \cite{Kim_4DSYM}. The moduli space of all such theories is connected. In one corner of the moduli space, we can think of this theory as the Heterotic theory on $S^1$ and in a different corner, it is the Type I theory on $S^1$. If we take a limit where the coupling of the Heterotic string goes to zero, the Dabholkar-Harvey BPS states will become light. On the other hand, if we keep that string coupling fixed, but take the radius to infinite distance, we will get light winding or KK states which are again BPS. On the other hand, if we take the type I theory on a circle and take the radius to infinity, we again get light KK states that are BPS. However, if we take the radius to zero, the T-dual description (type I' theory) can have no light BPS particles. As we will see, this corner of the 9d supergravity moduli space is special and we use Swampland arguments to provide a bottom-up explanation for its existence without relying on string theory. From this point onward, unless noted otherwise, type I, type II, or Heterotic refer to the supergravities and do not assume a string theory origin for them.

\subsection{Non-BPS limits}\label{NBPS}

In this work, we are  considering theories in 9 dimensions with 16 supercharges. We are particularly interested in investigating so-called ``non-BPS limits'' of 9d theories, which we defined below,  in order to establish the SLP in 9d.

\begin{framed}
A \textit{non-BPS limit} is defined as an infinite-distance limit in the moduli space in which there are no light BPS particles compared to the Planck scale.  
\end{framed}
As we will see, the non-BPS limits are very useful in identifying the emergent description.  In particular, we show that almost every such limit in 9d, decompactifies to a type I' background, which is a type IIA supergravity on an interval with a non-trivial profile of the dilaton with some BPS domain walls. 

\subsection{Moduli representation} \label{sec:nonbps}

As we outlined in the introduction, we are interested in infinite distance limits of theories with 16 supercharges where no BPS particles can become light. In what follows, we will assume the completeness of spectrum for BPS particles (which is indeed well-motivated in theories with 16 supercharges \cite{Bedroya21}).

Let us start with reviewing the bosonic content of theories with 16 supercharges in $d$ dimensions\footnote{The only exception is the chiral 6d theory which has tensor multiplets.}. 
\begin{itemize}
    \item Supergravity multiplet: $g_{\mu\nu}, B_{\mu\nu},A^i_\mu,\phi$ where $\phi$ is the dilaton and $i\in\{1,\hdots,10-d\}$.
    \item Vector multiplet: $A^\mu$ and $\Phi^j$ where $j\in\{1,\hdots,10-d\}$.
\end{itemize}
To see the structure of the multiplets, it is instructive to look at the toroidal compactification of a 10d Abelian theory with $r_{10}$ vector multiplets down to $d$ dimensions. This compactification will give us a theory with $r=r_{10}+(10-d)$ vector multiplets. This is because the 10d gravity multiplet breaks into a lower dimensional gravity multiplet and $10-d$ vector multiplets. 

In the following, we will identify which infinite distance limits are non-BPS limits. For that, let us start by considering the infinite distance limits for the scalars in the vector multiplets. 

The scalars of the vector multiplets have a scalar geometry\footnote{We ignore discrete quotients since our arguments will not be sensitive to that information about the global structure of the gauge group.} 
\begin{align} \mathcal{M}_V := \frac{ O(10-d,r) 
}{ O(10-d)\times O(r) }. \end{align}
This simply follows from supersymmetry \cite{Hohm:2014sxa}. We can arrange these scalars in a $10-d+r$ dimensional symmetric matrix $M$ such that 
\begin{align}\label{MC}
    M\begin{pmatrix}  \mathbb{I}_{10-d}& 0\\ 0 & -\mathbb{I}_{r}\\
    \end{pmatrix}M=\begin{pmatrix} \mathbb{I}_{10-d} & 0\\ 0 & -\mathbb{I}_{r}\\
    \end{pmatrix}.
\end{align}
The kinetic term for the gauge field is 
\begin{align}\label{SFC}
    -\frac{1}{4}e^{-2\phi}F^a_{\mu\nu}\left[ \begin{pmatrix}  \mathbb{I}_{10-d}& 0\\ 0 & -\mathbb{I}_{r}\\
    \end{pmatrix}M\begin{pmatrix}  \mathbb{I}_{10-d}& 0\\ 0 & -\mathbb{I}_{r}\\
    \end{pmatrix} \right]_{ab}F^b_{\mu\nu},
\end{align}
where $\phi$ is the dilaton in the gravity multiplet. This means that the gauge couplings are controlled by the entries of the matrix $M$. Since the gauge coupling also affect the gauge charges, the charge lattice depends on $M$ as well. The 16 supercharges almost fix the dependence of the charge lattice on the moduli (see Appendix D in \cite{Bedroya:2023}). The following transformation of $M$
\begin{align}\label{MT}
    M\rightarrow \Omega M\Omega ^T;~\Omega\in O(10-d,r),
\end{align}
transforms a vector $v$ in the charge lattice $\Lambda$ as 
\begin{align}
    v\rightarrow\Omega v.
\end{align}
If we identify the first $10-d$ entries of $v$ as $\vec{Q}_R$ and the rest with $\vec{Q}_L$, then the quantity $Q_{R}^2-Q_{L}^2$ is invariant. Moreover, it is easy to show that every transformation of $M$ that maintains \eqref{MC} can be expressed in the form \eqref{MT}. Therefore, we find that $Q_{R}^2-Q_{L}^2$ is invariant across the moduli space for any charge state. 

The infinite distance limits that are global geodesics, correspond to $\lim_{t\rightarrow\infty}\Omega(\gamma)M\Omega(\gamma)^T$ where $\Omega(\gamma)=\exp(\gamma U)$ is a boost in $(10-d,r)$. By global geodesic, we refer to geodesics that are the shortest path between any two points on them\footnote{For more on the definition of the global geodesic see Sec. \ref{sec:stringorkk}.}. Note that adding a rotation piece to the boost will lead to spiraling trajectories in the moduli space that are locally geodesic, but not globally. It is easy to show that an infinite boost always takes a null line to the origin.
\begin{figure}[H]
    \centering

\tikzset{every picture/.style={line width=0.75pt}} 

\begin{tikzpicture}[x=0.75pt,y=0.75pt,yscale=-1,xscale=1]

\draw  (178.5,150.15) -- (432.5,150.15)(303.63,30) -- (303.63,267) (425.5,145.15) -- (432.5,150.15) -- (425.5,155.15) (298.63,37) -- (303.63,30) -- (308.63,37)  ;
\draw [color={rgb, 255:red, 74; green, 144; blue, 226 }  ,draw opacity=1 ]   (206.5,53) -- (303.63,150.15) ;
\draw [color={rgb, 255:red, 74; green, 144; blue, 226 }  ,draw opacity=1 ]   (303.63,150.15) -- (400.77,247.31) ;
\draw [color={rgb, 255:red, 208; green, 2; blue, 27 }  ,draw opacity=1 ]   (303.63,150.15) -- (209.13,261.65) ;
\draw [color={rgb, 255:red, 208; green, 2; blue, 27 }  ,draw opacity=1 ]   (398.13,38.65) -- (303.63,150.15) ;
\draw    (250.21,224.64) .. controls (287.42,173.75) and (317.76,165.92) .. (367.5,226) ;
\draw [shift={(248.5,227)}, rotate = 305.64] [color={rgb, 255:red, 0; green, 0; blue, 0 }  ][line width=0.75]    (10.93,-3.29) .. controls (6.95,-1.4) and (3.31,-0.3) .. (0,0) .. controls (3.31,0.3) and (6.95,1.4) .. (10.93,3.29)   ;
\draw    (236.5,74) .. controls (303.17,148.63) and (315.38,131.18) .. (361.8,69.93) ;
\draw [shift={(362.5,69)}, rotate = 127.16] [color={rgb, 255:red, 0; green, 0; blue, 0 }  ][line width=0.75]    (10.93,-3.29) .. controls (6.95,-1.4) and (3.31,-0.3) .. (0,0) .. controls (3.31,0.3) and (6.95,1.4) .. (10.93,3.29)   ;

\draw (293,7.4) node [anchor=north west][inner sep=0.75pt]    {$Q_{L}$};
\draw (446,141.4) node [anchor=north west][inner sep=0.75pt]    {$Q_{R}$};

\end{tikzpicture}
    \caption{In any infinite boost of the charge lattice, a line in the light-cone ($|Q_L|=|Q_R|$) goes to the origin while the other points go to infinity. The line that goes to the origin is colored blue.}
    \label{fig:my_label}
\end{figure}
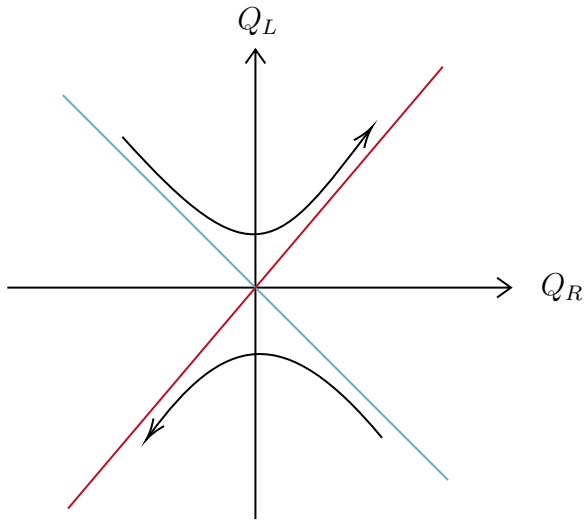

Therefore, for any infinite distance limit of $M$, a line of particles that satisfy $|Q_R|=|Q_L|$ and are in the plane of boost will go to the origin of the charge lattice. Let us call that line in the charge lattice, the \textit{ line of boost}.

The BPS condition in theories with 16 supercharges is \cite{Sen:1994eb}
\begin{align}
    m=|Q_R|.
\end{align}
Therefore, if there is a BPS particle with $|Q_R|=|Q_L|$ on the line of boost, that particle will become light in the infinite distance limit. However, we need to make sure that the two conditions $m=|Q_R|$ and $|Q_R|=|Q_L|$ can be simultaneously satisfied. For that, we use the BPS completeness hypothesis \cite{Kim:2019vuc,Hamada_recon} which states that in theories with 16 supercharges, any charge that is allowed to be BPS, is occupied by a BPS particle. To see what particles are allowed to be BPS, let us look at particles that are massive enough to be black holes. In that case, their mass must satisfy the extremality condition \cite{Sen:1994eb}.
\begin{align}
    m\geq |Q_L|.
\end{align}
The above inequality follows from the classical extremality condition which can receive higher derivative corrections \cite{Kats:2006xp} or quantum corrections \cite{Abedi:2017vtr}.
\begin{align}
    m\geq |Q_L|+\mathcal{O}(m_P^2/M)_\text{corrections}.
\end{align}
As long as the corrections on the right side are non-positive, the a BPS particle with $m=|Q_R|=|Q_L|$ is allowed. But the non-positivity of the corrections is guaranteed by the mild version of the Weak Gravity Conjecture (WGC) \cite{Arkani-Hamed:2006emk}. Therefore, based on WGC and BPS completeness, we argue that for any point of the charge lattice on the line of boost, at least one BPS particle or black hole of that charge exists. It could be the case that the line of boost does not intersect with the lattice at all. In that case, all the BPS particles become massive and the corresponding limit is non-BPS. However, if the line of boost intersects with a BPS particle, the BPS particles at the intersection will become massless at the infinite distance limit. 

Let us make the above discussion more explicit. Consider the following infinite distance limit generated by a boost.

\begin{align}\label{GGP} \Omega_{a,i}(\gamma) = \mt{ e_{aa}\cosh \gamma  & e_{ai}\sinh \gamma \\  e_{ia}\sinh\gamma & e_{ii}\cosh\gamma },  \end{align}
where $e_{pq}$ is a matrix with one non-zero element of $1$ in row $p$ and column $q$. The the two axes acted upon are $a \in \{1, \cdots, 10-d\}$ axis and $i \in \{1, \cdots, r\}$ axis. 

An infinite-distance limit in the moduli space therefore corresponds to $\g \to \pm \infty$. Now, take the action of $\Omega_{ai}$ on the charge vector $\V{p} = (Q_R, Q_L)$. 

Under the action of $\Omega_{a,i}$ on $M$, a charge vector $v = (Q_R, Q_L)$ goes to 
\begin{align} |Q_{R, a}| \mapsto |Q_{R, a} \cosh\g + Q_{L, i} \sinh\g| = |\frac{1}{2} (Q_{R, a} + Q_{L, i}) e^\g + \frac{1}{2} (Q_{R, a} - Q_{L, i}) e^{-\g} | \end{align}

Depending on the sign of $\gamma$ in the infinite distance limit, the line of boost is \begin{align}
    Q_{R,b}&\propto \delta_{ba}\nonumber\\
    Q_{L,j}&\propto \delta_{ji}\nonumber\\
    Q_{L,i}&=\pm Q_{R,a}.
\end{align} 
Suppose we have BPS particles on both lines of boosts. As we take $\gamma \to \pm\infty$,  using $m=|Q_R|$ we find that $m \sim e^{|\g|}$, unless we have $Q_{R, a} = \mp Q_{L, i}$, in which case the masses scale as $m \sim e^{\mp \g}$. Based on the sign of $\gamma$ such particles become massless in the limit $\gamma\rightarrow +\infty$ or $\gamma\rightarrow -\infty$. So in the presence of BPS particles on lines of boosts, any infinite distance limit in the moduli space of the vector multiplets could yield light BPS particles.  

However, as one can see from \eqref{SFC}, the gauge couplings also depend on the dilaton. Therefore, the gauge charges (and hence masses) of the BPS particles also depend on the dilaton as $m\propto e^{\phi}$. However, the expression in \eqref{SFC} is in the string frame. After going to the Einstein frame, we find the following mass relation in terms of the canonically normalized dilaton $\hat\phi$.
\begin{align}\label{BPSM} m \simeq e^{\pm \g_{a, i}} e^{\frac{d}{2\sqrt{d-2}}\hat\phi} m_p. \end{align}

We conclude that even if there are BPS particles on the lines of boost, all of them will become heavy in the limit $\gamma\to \pm \infty$ if we also take $\hat\phi$ to infinity such that $\abs{\hat\phi}/|\gamma| \geq (2\sqrt{d-2}/d)$. Note that in these limits, there is no light BPS string.

To conclude, we showed that any infinite distance limit $|\gamma|,\hat\phi\rightarrow\infty$ that $\abs{\hat\phi}/|\gamma| \geq (2\sqrt{d-2}/d)$ is a non-BPS limit.  

\subsection{Tensionless string or KK tower?} \label{sec:stringorkk}

In this section, we will use the sharpened distance conjecture to argue that there are only countably many infinite-distance limits in which the leading light tower is a string tower. In particular, all but countably many infinite distance limits decompactify. As a consequence, we find that if the leading tower in a non-BPS limit is a string tower; any infinite distance limit sufficiently close to it will decompactify. Our argument has two steps: 1) the number of inequivalent strings is countable. 2) Every string limit has a unique inequivalent (up to dualities) direction in the moduli space. 
\vspace{10pt}

\noindent\textbf{Step 1: countable strings}
\vspace{5pt}

We use the finiteness of black hole entropy to argue that the number of inequivalent strings is countable. Suppose, for the sake of a contradiction, that there are uncountably many inequivalent strings each of which has a worldsheet theory which is trustable in some region of the moduli space $\mathcal{M}$. We will argue that this contradicts the black hole entropy bound at some point in the moduli space. 

Suppose we have uncountably many string limits labelled by $\alpha\in I$ where $I$ is uncountable. Consider an arbitrarily small positive number $\epsilon$ such that for each string limit, there is a distinct point $p_\alpha$ in the moduli space such that the corresponding string description is valid in a neighborhood of radius $\epsilon$ around $p_\alpha$. We can partition $\mathcal{M}$ into countably many neighborhoods with diameter smaller than $\epsilon$. Let us explain why this is possible. 

For a given point $p$, we can write the moduli space as the countable union of $\cup_{n\in\mathds{N}}D_n$ where $D_n$ is a closed disc with radius $n/\kappa$ with respect to the canonical metric. Assuming the moduli space is completely regular and Hausdorff, the closed discs admit Stone–\v{C}ech compactification $\bar{D}_n$. Since $\bar{D}_n$ are compact and have finite volume, we can cover them with a finite number of neighborhoods each with a diameter less than $\epsilon$. Therefore, we can cover the moduli space $\mathcal{M}\subset \cup_n\bar{D}_n$ with a countable number of neighborhoods of diameters less than $\epsilon$. Let us call these neighborhoods $U_n$ where $n$ is a positive integer.

 Now that the moduli space is covered by countably many neighborhoods $U_n$, one of them must include uncountably many of the points $\{p_\alpha|\alpha\in I\}$. Let us call that neighborhood $U$. Therefore, there are uncountably many distinct strings in $U$. Now we show that there exists a finite cutoff $\Lambda_c$, such that uncountably many of the strings have mass scales below $\Lambda_c$.

Suppose $V_n$ is the subset of the points $p_\alpha$ in $U$ such that the mass scale of the string $\alpha$ is less than $n m_P$ in $U$. Since $\cup_{n\in \mathds{N}} V_n$ contains all the uncountably many strings in $U$, for some $m$, the set $V_m$ must be uncountable. Therefore, there are uncountably many strings in the neighborhood $U$ that have mass scale below $\Lambda_c=m\cdot m_P$. However, this violates the finiteness of the Bekenstein-Hawking entropy, and therefore must not be allowed. Therefore, our original assumption is incorrect, and the number of inequivalent strings must be countable. Note that our argument so far has not relied on the strings being weakly coupled. However, that would be important for the second step of the argument.

\vspace{10pt}

\noindent\textbf{Step 2: rigidity of string limits}
\vspace{5pt}

In this step, we want to show that each one of the countable many string limits forms the leading tower in most one direction in the moduli space. We show that if one makes a small change in the direction of the infinite distance limit away from a string limit, the leading tower is no longer a string tower. But first, let us define a useful notion to study the infinite distance limits. We define a global geodesic in the moduli space to be a geodesic which is the shortest path between any two points on it. We can think of infinite distance limits as global geodesics. This refinement excludes infinite geodesics that are stuck in a compact subspace. Note that not every infinite geodesic that starts at a given point of the moduli space is a global geodesic. For example, in type IIB string theory, for any point $p$ in the moduli space, there is a unique global geodesic that starts at $p$. This is because, for any two BPS strings, the limits where they become tensionless are mapped to each other via duality. Changing the angle of the initial velocity of the geodesic will create a geodesic which either does not get to infinity or winds around the fundamental domain of $SL(2,\mathds{Z})$ and is not a global geodesic. Therefore, even if the new path is locally a geodesic, it is not a global geodesic. Due to this, we would say the type IIB theory has a unique inequivalent infinite distance limit. In the rest of this section, we will show that if we make a sufficiently small change in the direction of a string limit, the resulting limit cannot be a string limit. We assume that such a sufficiently small variation can change one global geodesic to another. Otherwise, if the limit is rigid, then we trivially know that there is no nearby string limits. 

Sharpened distance conjecture tells us that in a string limit, the tension of the string will go like 
\begin{align}\label{dccs}
\sqrt{T}\propto m_p e^{-{\kappa\over \sqrt{d-2}}\hat\phi},
\end{align}
where $\hat\phi$ is the canonically normalized distance in the field space. If we make a sufficiently small change in the direction of the infinite distance direction that combines the modulus $\hat\phi$ with another spacetime modulus, the distance travelled in the moduli space per change of $\hat\phi$ will increase\footnote{We assume that $\hat\phi$ is a linear spacetime modulus as opposed to something like $\hat\phi=\sqrt{\phi_1^2+\phi_2^2}$ which would give the same rate of change of $\phi$ in every infinite distance limit in $(\phi_1,\phi_2)$ plane. By linear modulus, we mean that if a global geodesic at point $p$, which is in the asymptotics of the moduli space, has an infinitesimal angular separation $\delta \theta$ from $\partial_{\hat\phi}$ then $\Delta \hat\phi/\Delta l$ converges to $\cos(\theta)$ where $\Delta l$ is the canonical distance. Note that this follows from the emergent string conjecture which conjectures that the string is fundamental. In that case, the modulus $\phi$ can be shown to always be a linear spacetime modulus. This follows as a corollary of the argument for the sharpened distance conjecture from the emergent string conjecture in \cite{Agmon:2022thq}.}. This would lead to a decrease in the coefficient $1/\sqrt{d-2}$ in the exponent in \eqref{dccs}. Therefore, the sharpened distance conjecture implies that the string \eqref{dccs} can no longer be the leading tower anymore. So we find that any worldsheet theory has at most one infinite distance limit where the string is the leading tower which corresponds to the direction in which the tension decays the fastest.

By putting steps 1 and 2 together, we conclude that there are only countably many infinite distance limits in which the leading tower is a string tower. This is a powerful result because it means that even though the direction we have considered has a tensionless string as the lightest tower, the nearby infinite distance limits must be decompactification limits. For our purposes, this will mean that if there is a non-BPS limit with a string leading tower, any sufficiently close non-BPS limit decompactifies.

\section{Non-BPS limits in 9d}

In this section, we study the non-BPS limits in 9d theories that decompactify, which as we discussed in the previous sections cover almost all of the directions. Here, by \textit{decompactification}, we mean that the lightest tower of states can be obtained from a dimensional reduction of a higher dimensional supergravity background. In Sec. \ref{subsec:classlim}, we will explicitly classify the decompactification limits of the 9d theory. We will find that non-BPS limits necessarily arise in the so-called \textit{Type I' corner} (which is the massive IIA supergravity background on an interval). In Sec. \ref{sec:typei'sugra}, we will briefly define and review the features of type I' supergravity and describe the relevant moduli. Finally, in Sec. \ref{sec:matching}, we will put together all of the pieces and describe a matching between the 9d theory and type I' moduli in the non-BPS limits. 

\subsection{Classification of the limits} \label{subsec:classlim}

Let us first classify the possible decompactification limits of the 9d theory. The 9d theory can only decompactify to either a 10d or an 11d which is the highest dimension for supergravity. Therefore, there are two cases to consider.  

\subsubsection*{11d Backgrounds}

We firstly consider the decompactification to 11d supergravity, which is unique at low energies \cite{Cremmer:1978km,Freedman:2012zz}. Therefore, different decompactification limits to 11d correspond to different choices of compact dimensions and boundary conditions if the compact directions have any boundaries. Since we wish to preserve 16 supercharges, the choices of the compact directions are fairly limited. The only boundary-less smooth two-dimensional Ricci-flat manifolds are torus Klein bottle, and the ones with at most one non-trivial cycles are M\"{o}bius strip or cylinder. 

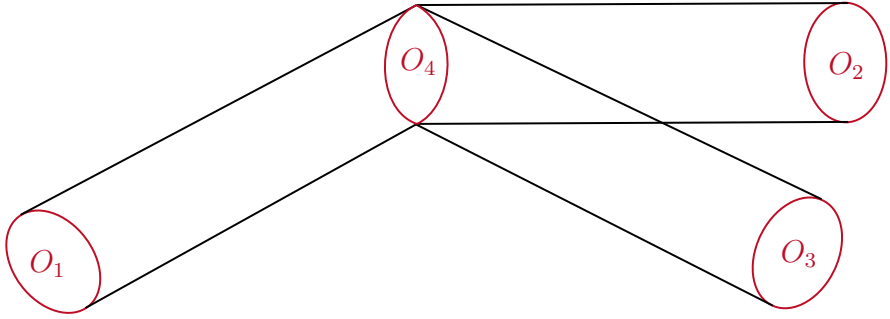
\begin{figure}
    \centering

\tikzset{every picture/.style={line width=0.75pt}} 

\begin{tikzpicture}[x=0.75pt,y=0.75pt,yscale=-1,xscale=1]

\draw  [color={rgb, 255:red, 208; green, 2; blue, 27 }  ,draw opacity=1 ] (521.35,65.49) .. controls (513.87,52.76) and (514.81,33.71) .. (523.45,22.94) .. controls (532.1,12.17) and (545.17,13.77) .. (552.65,26.51) .. controls (560.13,39.24) and (559.19,58.29) .. (550.55,69.06) .. controls (541.9,79.83) and (528.83,78.23) .. (521.35,65.49) -- cycle ;
\draw    (320.79,17) -- (538.5,16) ;
\draw    (320,77) -- (538.5,76) ;
\draw  [color={rgb, 255:red, 208; green, 2; blue, 27 }  ,draw opacity=1 ] (490.46,152.88) .. controls (489.02,138.63) and (497.88,122.26) .. (510.24,116.3) .. controls (522.6,110.34) and (533.78,117.06) .. (535.22,131.31) .. controls (536.66,145.55) and (527.81,161.93) .. (515.45,167.88) .. controls (503.09,173.84) and (491.9,167.12) .. (490.46,152.88) -- cycle ;
\draw    (320.79,17) -- (525.08,115.01) ;
\draw    (320,77) -- (500.43,168.72) ;
\draw  [color={rgb, 255:red, 208; green, 2; blue, 27 }  ,draw opacity=1 ] (134.07,169.45) .. controls (121.13,162.85) and (112.21,147.25) .. (114.15,134.59) .. controls (116.09,121.94) and (128.14,117.04) .. (141.08,123.64) .. controls (154.01,130.24) and (162.93,145.84) .. (160.99,158.49) .. controls (159.06,171.14) and (147,176.05) .. (134.07,169.45) -- cycle ;
\draw    (121.11,122.86) -- (320.79,17) ;
\draw    (153.75,169.94) -- (320.91,77.03) ;
\draw [color={rgb, 255:red, 208; green, 2; blue, 27 }  ,draw opacity=1 ]   (320.79,17) .. controls (344.5,31) and (338.5,69) .. (320.91,77.03) ;
\draw [color={rgb, 255:red, 208; green, 2; blue, 27 }  ,draw opacity=1 ]   (320.79,17) .. controls (300.5,26) and (298.5,69) .. (320.91,77.03) ;

\draw (124,139.4) node [anchor=north west][inner sep=0.75pt]  [color={rgb, 255:red, 208; green, 2; blue, 27 }  ,opacity=1 ]  {$O_{1}$};
\draw (527,39.4) node [anchor=north west][inner sep=0.75pt]  [color={rgb, 255:red, 208; green, 2; blue, 27 }  ,opacity=1 ]  {$O_{2}$};
\draw (503,134.4) node [anchor=north west][inner sep=0.75pt]  [color={rgb, 255:red, 208; green, 2; blue, 27 }  ,opacity=1 ]  {$O_{3}$};
\draw (311,37.4) node [anchor=north west][inner sep=0.75pt]  [color={rgb, 255:red, 208; green, 2; blue, 27 }  ,opacity=1 ]  {$O_{4}$};

\end{tikzpicture}
    \caption{An exotic 11d supergravity background on a 2d space resulted from identifying a boundary of three cylinders. The existence of such a background depends on the existence on the appropriate boundary theories $O_i$, which is a UV question. For example, there is no evidence for the existence of such boundary theories in string theory.}
    \label{EM}
\end{figure}

However, one can \textit{a priori} imagine smooth manifolds with more boundaries. We could also consider singular combinations of such backgrounds by gluing multiple components along the boundaries with some degrees of freedom living on the boundary (for example as in Figure \ref{EM}). However, there is no evidence of such exotic backgrounds in string theory. This question should be addressed also beyond string theory, since whether such a boundary theory exists is a UV question. We provide a bottom-up argument based on the classification of 5d SCFTs for why such exotic backgrounds do not exist. 

Let us consider the small instantons of the 9d theory (for definition see \ref{sec:gaugesymm}). Using the BPS completeness hypothesis \cite{Kim:2019vuc}, we assume BPS small instantons exist. Furthermore, the strong version of the cobordism conjecture in \cite{McNamara:2019rup} implies that the moduli space of these small instantons is connected. Now, a generic BPS 4-brane in 9d supergravity will carry a 5-form charge\footnote{We will give a detailed argument for this point in \ref{sec:gaugesymm}, but is a summary of the argument. The number of gauge instantons and branes with 5-brane charge is not independently conserved due to gauge symmetry. Therefore, they must be transformable to each other and share a moduli space.}. However, the 5-form gauge potential in 9d must be the KK reduction of the dual of the 3-form gauge potential in 11d. Therefore, the 9d small instantons must correspond to wrapped BPS 5-branes in 11d supergravity picture. When the 5-brane approchaes the boundary, the small instanton will approach the boundary of its moduli space.

We note first that the compact dimension cannot be disconnected, as if it were, we would find two copies of the supergravity multiplet in lower dimension. Therefore, the uniqueness of the lower dimensional graviton implies that the compact dimension is connected.

Now, from the rank 1 classification of 5d SCFTs, we know that any local piece of the Coulomb branch of an SCFT or a free theory must either look like an open interval which has at most one closed end \cite{Seiberg:1996bd,Douglas:1996xp,Morrison:1996xf,Bhardwaj:2019jtr,Bhardwaj:2019fzv}. The one dimensional Coulomb branch of an SCFT has the local structure of $\R/\Z_2$ \cite{Seiberg:1996bd} but not a star shaped singularity. In other words, star-shaped moduli spaces are not allowed. Thus, the compact dimensions (corresponding to the moduli space of a 5d theory in its IR) can have at most two boundaries, and the only singularities/boundaries allowed in the moduli space of 5d theories (either free or SCFT) are $\mathbb{R}/\mathbb{Z}_2$.

Note that compactifying the theory on a torus will result in a theory with 32 supercharges rather than 16. Thus, in summary, we get the following possibilities for each piece of the compact dimension based on supersymmetry:
\begin{itemize}
    \item Cylinder
    \item M\"{o}bius strip 
    \item Klein bottle
\end{itemize}

The only backgrounds that are supported by string theory via dualities are those where the internal geometry is exactly one of the above possibilities, and the boundary theory is that of the Ho\v{r}ava-Witten wall \cite{Horava96}.

Note that all of the above manifolds have an $S^1$ piece with a $U(1)$ symmetry which further increases the rank by 1. Most importantly, the KK tower along the $S^1$ piece will be a light BPS tower. Therefore, no decompactification to 11d supergravity (even the exotic ones suppose they exist) is a non-BPS limit according to our definition from the previous section.

\subsubsection*{10d Backgrounds}

Suppose our theory decompactifies to a 10d theory. If the local 10d EFT has 16 supercharges, the compact dimension needs to preserve all of that supersymmetry. The only option for such a compact dimension is $S^1$. However, just like in the previous case, having an $S^1$ as a compact dimension ensures that there is a BPS KK tower in the decompactification limit. Therefore, such a limit is not a non-BPS limit. 

The other possibility is that the local physics is described by a supergravity with 32 supercharges, but the global structure of the compact dimension breaks half the supersymmetry. For example, at the massless level, we can have accidental supersymmetry which is broken by the massive states. Such 10d low-energy effective field theories can type II supergravity theory on $S^1$ or on a 1d manifold with boundaries. If the theory decompactifes to a type II supergravity on $S^1$, the KK particles with even charges will form a light BPS tower in the 9d theory. Therefore, such a limit would not be a non-BPS limit. Now let us consider type II supergravities on 1d manifolds with boundaries. Since the type IIB supergravity is chiral, putting it on a manifold with a co-dimension 1 boundary would break all supersymmetry, whether or not they are accidental.

Type IIA supergravity however allows BPS end of the universe walls and BPS domain walls since it is non-chiral. Therefore, the theory can decompactify to  a type IIA background on a union of circles and intervals that are joined via some boundary defects and some BPS domain walls on the inside. Even though the supergravity allows for such a background, their existence depends on the spectrum of the non-perturbative 8-branes which is a UV information. In the following, we use the classification of 5d SCFTs to argue that the only such type IIA background that could exist is type IIA on an interval.

\begin{figure}[H]
    \centering

\tikzset{every picture/.style={line width=0.75pt}} 

\begin{tikzpicture}[x=0.75pt,y=0.75pt,yscale=-1,xscale=1]

\draw    (219,22) -- (343.5,101) ;
\draw    (223.5,177) -- (343.5,101) ;
\draw    (343.5,101) .. controls (362.5,55) and (399.8,50.7) .. (412.5,83) .. controls (425.2,115.3) and (380.5,186) .. (343.5,101) -- cycle ;
\draw  [fill={rgb, 255:red, 208; green, 2; blue, 27 }  ,fill opacity=1 ] (340.25,100.25) .. controls (340.25,98.46) and (341.71,97) .. (343.5,97) .. controls (345.29,97) and (346.75,98.46) .. (346.75,100.25) .. controls (346.75,102.04) and (345.29,103.5) .. (343.5,103.5) .. controls (341.71,103.5) and (340.25,102.04) .. (340.25,100.25) -- cycle ;
\draw  [fill={rgb, 255:red, 208; green, 2; blue, 27 }  ,fill opacity=1 ] (215.75,22) .. controls (215.75,20.21) and (217.21,18.75) .. (219,18.75) .. controls (220.79,18.75) and (222.25,20.21) .. (222.25,22) .. controls (222.25,23.79) and (220.79,25.25) .. (219,25.25) .. controls (217.21,25.25) and (215.75,23.79) .. (215.75,22) -- cycle ;
\draw  [fill={rgb, 255:red, 208; green, 2; blue, 27 }  ,fill opacity=1 ] (220.25,177) .. controls (220.25,175.21) and (221.71,173.75) .. (223.5,173.75) .. controls (225.29,173.75) and (226.75,175.21) .. (226.75,177) .. controls (226.75,178.79) and (225.29,180.25) .. (223.5,180.25) .. controls (221.71,180.25) and (220.25,178.79) .. (220.25,177) -- cycle ;
\draw  [fill={rgb, 255:red, 74; green, 144; blue, 226 }  ,fill opacity=1 ] (234.75,34) .. controls (234.75,32.21) and (236.21,30.75) .. (238,30.75) .. controls (239.79,30.75) and (241.25,32.21) .. (241.25,34) .. controls (241.25,35.79) and (239.79,37.25) .. (238,37.25) .. controls (236.21,37.25) and (234.75,35.79) .. (234.75,34) -- cycle ;
\draw  [fill={rgb, 255:red, 74; green, 144; blue, 226 }  ,fill opacity=1 ] (253.75,46) .. controls (253.75,44.21) and (255.21,42.75) .. (257,42.75) .. controls (258.79,42.75) and (260.25,44.21) .. (260.25,46) .. controls (260.25,47.79) and (258.79,49.25) .. (257,49.25) .. controls (255.21,49.25) and (253.75,47.79) .. (253.75,46) -- cycle ;
\draw  [fill={rgb, 255:red, 74; green, 144; blue, 226 }  ,fill opacity=1 ] (290.75,70) .. controls (290.75,68.21) and (292.21,66.75) .. (294,66.75) .. controls (295.79,66.75) and (297.25,68.21) .. (297.25,70) .. controls (297.25,71.79) and (295.79,73.25) .. (294,73.25) .. controls (292.21,73.25) and (290.75,71.79) .. (290.75,70) -- cycle ;
\draw  [fill={rgb, 255:red, 74; green, 144; blue, 226 }  ,fill opacity=1 ] (309.75,82) .. controls (309.75,80.21) and (311.21,78.75) .. (313,78.75) .. controls (314.79,78.75) and (316.25,80.21) .. (316.25,82) .. controls (316.25,83.79) and (314.79,85.25) .. (313,85.25) .. controls (311.21,85.25) and (309.75,83.79) .. (309.75,82) -- cycle ;
\draw  [fill={rgb, 255:red, 74; green, 144; blue, 226 }  ,fill opacity=1 ] (251.75,158) .. controls (251.75,156.21) and (253.21,154.75) .. (255,154.75) .. controls (256.79,154.75) and (258.25,156.21) .. (258.25,158) .. controls (258.25,159.79) and (256.79,161.25) .. (255,161.25) .. controls (253.21,161.25) and (251.75,159.79) .. (251.75,158) -- cycle ;
\draw  [fill={rgb, 255:red, 74; green, 144; blue, 226 }  ,fill opacity=1 ] (403.75,125) .. controls (403.75,123.21) and (405.21,121.75) .. (407,121.75) .. controls (408.79,121.75) and (410.25,123.21) .. (410.25,125) .. controls (410.25,126.79) and (408.79,128.25) .. (407,128.25) .. controls (405.21,128.25) and (403.75,126.79) .. (403.75,125) -- cycle ;
\draw  [fill={rgb, 255:red, 74; green, 144; blue, 226 }  ,fill opacity=1 ] (400.75,71) .. controls (400.75,69.21) and (402.21,67.75) .. (404,67.75) .. controls (405.79,67.75) and (407.25,69.21) .. (407.25,71) .. controls (407.25,72.79) and (405.79,74.25) .. (404,74.25) .. controls (402.21,74.25) and (400.75,72.79) .. (400.75,71) -- cycle ;

\end{tikzpicture}
    \caption{The internal geometry for an exotic type IIA background. The red points are boundary theories and the blue points represent BPS domain walls. We provide a bottom-up argument from the classification of 5d SCFTs that rules out such backgrounds.}
    \label{SSGA}
\end{figure}
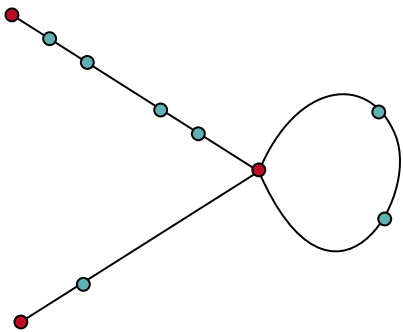

In section \ref{sec:gaugesymm}, we use the no global symmetry conjecture to show that BPS 4-branes in such type IIA backgrounds are small instantons in 9d which have a rank 1  worldvolume theory with 8 supercharges. We momentarily borrow that fact to show that exotic type IIA backgrounds such as \ref{SSGA} do not exist. A 4-brane probing a defect at a star-shaped node such as that in \ref{SSGA} leads to a star-shaped Coulomb branch for a rank 1, 5d $\mathcal{N}=1$ SCFT or free theory which describes the low-energy physics of the 4-brane. 

On the other hand, from the rank 1 classification of 5d SCFTs, we know that any local piece of the Coulomb branch of an SCFT or a free theory must either look like an open interval which has at most one closed end \cite{Seiberg:1996bd,Douglas:1996xp,Morrison:1996xf,Bhardwaj:2019jtr,Bhardwaj:2019fzv}. The one dimensional Coulomb branch can have a local singular structure of $\R/\Z_2$ \cite{Seiberg:1996bd} but not a star-shaped singularity with more than two legs. In other words, star-shaped moduli spaces are not allowed. Therefore, the only candidate for a type  IIA background with a boundary is type IIA supergravity on an interval. Note that a disjoint union of intervals is not allowed as it leads to multiple gravitons in the lower dimensional theory. 

Before continuing, we note that from the classification of the 5d SCFTs, we can also identify when the dilaton must diverge at the endpoints of the interval. If the 4-brane theory probing the endpoint of the interval is described by a free theory at low energies, the dilaton does not diverge. However, if the theory is a 5d SCFT, the dilaton must diverge. This is because the coupling of all 5d SCFTs diverge at the boundary of their Coulomb branch. Since this coupling is controlled by the spacetime dilaton, the dilaton must also diverge.

A massive type IIA supergravity background on an interval is nothing other than type I' supergravity. In the following subsection, we define and review the general properties of type I' supergravity backgrounds without relying on string theory.

\subsubsection*{Summary}

In summary, we find that there is only one non-BPS decompactification limit of the 9d theory -- the type I' supergarvity. In the rest of this section, we will review the properties of type I' supergravity and give an explicit matching between the moduli of 9d supergravity and type I'. We will find that the non-BPS decompactification limits can all be described by the type I' picture, and this will allow us to access information about the gauge symmetry of the theory in Sec \ref{sec:gaugesymm}.

\subsection{Type I' supergravity} \label{sec:typei'sugra}

In the previous subsection, we argued that there is only one non-BPS decompactification of the 9d theory corresponding to the  massive IIA supergravity background on an
interval or the Type I' description.
The Type I' theory is typically understood as a string theory but we would like to define it as a supergravity.  To that end, we will briefly review the massive type IIA supergravity. In particular, the type IIA supergravity has a mass deformation which adds a parameter known as the Romans mass \cite{Romans:1985}. The case where the Romans mass is set to zero reduces to the ordinary IIA supergravity. There are two important features to keep in mind about the massive type IIA supergravity.
\begin{itemize}
    \item the Romans mass is not a dynamical field and must be thought as a parameter of the theory\footnote{Note that the application of no global symmetry conjecture \cite{Heidenreich:2020pkc} to ($-1$)-form symmetry implies that no theory of quantum gravity must have a free parameter. For the massive type IIA, this implies that massive type IIA in 10 non-compact dimensions with an arbitrary value of the Romans mass belongs to the Swampland (see \cite{Aharony:2010af} for a string theory argument). Therefore, we expect the profile of the Romans mass be set by boundary conditions and defects. As we will shortly see, this is indeed the case in the backgrounds of massive IIA on an interval.}.
    \item Due to the coupling of the Romans mass to the dilaton, any non-zero value of the Romans mass creates a linear profile for $e^{-\phi}$ in space.
\end{itemize}
We would like to study the supersymmetric backgrounds of massive type IIA theory on an interval. For bottom-up reasons that we will explain in section \ref{sec:gaugesymm}, in the presence of $e_n\oplus e_m$ gauge algebra enhancements, $e^{-\phi}$ must vanish at the endpoints of the interval. Therefore, the Romans mass needs to change along the interval to allow for a profile of $e^{-\phi}$ that starts and ends at zero. For that we must allow for jumps in the Romans mass which are mediated by supersymmetric 8-branes. Such branes might or might not exist in the UV theory, but in this section, our analysis is limited to what is or is not allowed in field theory. So we postpone those concerns to future sections. 

The supersymmetric 8-branes that mediated the jump must satisfy a BPS condition involving a 9-form charge. In this case, the 9-form gauge field must be dual to the Romans mass. Therefore, the charges of the 8-branes and, consequently, the jumps in the Romans mass must be quantized\footnote{In string theory these are the D8 branes. But here we do not rely on string theory and therefore, so far, we do not have any bottom-up information on the worldvolume theory of one or a stack of the 8-branes.}. We take the tension of the 8-brane with a unit 9-form charge to be $\mu_8$.

\begin{statement*}
Type I' supergravity is a supersymmetric background of massive type IIA on an interval with a configuration of BPS domain walls perpendicular to the interval. The domain walls create a profile for the dilaton such that $e^{-\phi}$ vanishes at the endpoints of the interval. This is a field theory definition and the existence of such a background depends on the UV properties such as the spectrum of non-perturbative BPS domain walls and end-of-the-universe walls. 
\end{statement*}

Now we are ready to use the equations of the supergravity, to find the profile of the dilaton based on the positions of the supersymmetric 8-branes. The calculations parallel those carried in the context of string theory, however, we ignore any UV input and focus on the supergravity. 

Suppose we have 9 large dimensions and a single compact dimension parametrized by $x^9$ such that $0 \leq x^9 \leq 2\pi$ and $x^9 \sim -x^9$. There are end of the universe walls sitting at the endpoints $x^9 = 0$ and $x^9 = \pi$, which we do not know their microscopic description.

The type I' supergravity action is
\begin{align} \label{TypeI} S_{I'} = \int d^{10} x \sqrt{-g} e^{-2\phi_{10}} \left( \frac{1}{2} R + 2 \p_M \phi_{10} \p^M \phi_{10}  \right) - \frac{1}{2} \int F^* F + \cdots \end{align}
Here we have included only the bosonic terms coming from the metric and the top-form gauge field strength $F$ which is dual to the Romans mass. The BPS 8-brane solutions are electrically coupled to the 9-form gauge field $A$ with field strength $F$. Suppose the 8-branes are positioned at $x^9_i$. Let the indices $M, N \in \{0, \cdots, 9\}$, and let $\u, \nu \in \{0, \cdots, 8\}$. Suppose further that there are $n_1$ branes at $x_1^9$, $n_2$ branes at $x_2^9$, and so on.

 The equation of motion for the 9-form field $A$ yields
\begin{align} F = \nu_0 dx^0 \wedge \cdots \wedge dx^9 , \end{align}
where $\nu_0$ is a piecewise constant function which jumps at the positions of the branes. 

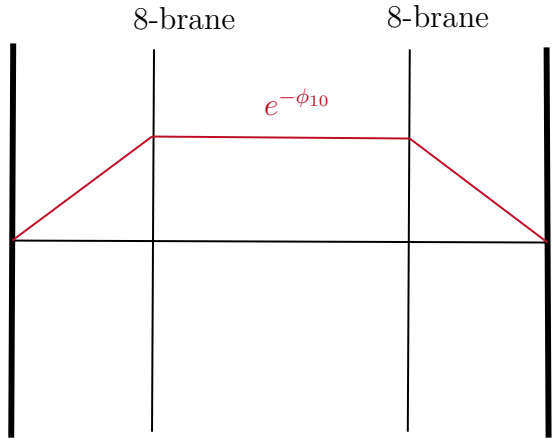
\begin{figure}
    \centering
    \tikzset{every picture/.style={line width=0.75pt}} 

\tikzset{every picture/.style={line width=0.75pt}} 

\begin{tikzpicture}[x=0.75pt,y=0.75pt,yscale=-1,xscale=1]

\draw [line width=2.25]    (120,55) -- (119,254) ;
\draw [line width=2.25]    (389,57) -- (390,254) ;
\draw    (119.5,154.5) -- (389.5,155.5) ;
\draw    (191,58) -- (190,251) ;
\draw    (320,58) -- (319,251) ;
\draw [color={rgb, 255:red, 208; green, 2; blue, 27 }  ,draw opacity=1 ]   (119.5,154.5) -- (190,102) ;
\draw [color={rgb, 255:red, 208; green, 2; blue, 27 }  ,draw opacity=1 ]   (320,103) -- (190,102) ;
\draw [color={rgb, 255:red, 208; green, 2; blue, 27 }  ,draw opacity=1 ]   (389.5,155.5) -- (320,103) ;

\draw (179,35.4) node [anchor=north west][inner sep=0.75pt]    {8-brane};
\draw (307,34.4) node [anchor=north west][inner sep=0.75pt]    {8-brane};
\draw (245,76.4) node [anchor=north west][inner sep=0.75pt]  [color={rgb, 255:red, 208; green, 2; blue, 27 }  ,opacity=1 ]  {$e^{-\phi_{10}}$};

\end{tikzpicture}
    \caption{A schematic illustration of type I' supergravity with two BPS end of the universe walls and two BPS 8-branes located at the specified positions. The red line describes the profile of the type I' dilaton (in the string frame instead of the Einstein frame) on the compact interval. Note also that the distance between the two 8-branes in the middle can be arbitrary.}
    \label{fig:typei'}
\end{figure}

Since the supergravity equations are identical to the ones appearing in type I' string theory, we can use the solution in \cite{Polchinski95}. However, due to lack of any bottom-up knowledge about the number of 8-branes at this point, we have to generalize that solution to an arbitrary number of 8-branes. The profile of the type I' dilaton and the type I' metric may be parametrized by a constant $C$ and a piecewise constant function $B(x^9)$ whose value changes at the position of each brane and which is entirely determined by the value $B(0)$. In conformal gauge, $g_{MN} = \Omega^2(x^9) \eta_{MN}$, and we then find 
\begin{align} e^{\phi_{10}(x^9)} = z(x^9)^{-5/6}, \quad & \Omega(x^9) = C z(x^9)^{-1/6} \\
z(x^9) = 3C ( B \mu_8 & - \nu_0 x^9)/\sqrt{2}. \end{align}
Here $\nu_0 = \nu_0(x^9)$ is a piecewise constant function with jump discontinuity $\Delta \nu_0 = n_i \mu_8$ at each stack of $n_i$ 8-branes. See Fig. \ref{fig:typei'} for an illustration of the type I' dilaton profile on the compact interval. 

\subsection{Matching 9d and type I' moduli} \label{sec:matching}

In this subsection, we will use the type I' supergravity action to express the 9d moduli discussed in section \ref{sec:nonbps}  in terms of the type I' parameters. Similar to the previous subsection, our study is  purely field theoretic and does not rely on string theory. For example, the number of branes are left arbitrary to be determined by the Swampland principles which capture the restrictions imposed by the consistency of a UV completion. Having said that, our calculations must parallel the existing results in string theory \cite{Polchinski95} when the number of branes match.  

The type I' action eq. \eqref{TypeI} is expected to be exact for BPS configuration where no light field (except the non-trivial profile of dilaton) is excited. Therefore, the dependence of the action on the position of the 8-branes as 9d scalar fields is protected by supersymmetry. This allows us to find the scalar kinetic terms of the 9d action and match them with the moduli in  eq. \eqref{TypeI}. Let us first start with the 9d dilaton $\phi$. If we dimensionally reduce the action \eqref{sec:nonbps} we find
\begin{align}
 e^{-2\phi} = \mu_8^\frac{7}{2} (2)^{-\frac{7}{4}} (4\pi)^7 C^{-10/3} \left( \int_0^{2\pi} dx^9 w(x^9) \right)^{-5/2},
\end{align}

where 
\begin{align} w(x^9) = 3^{1/3} 2^{-1/6}\left[ \mu_8 B(x^9) -  x^9 \nu_0(x^9) \right]^{1/3}. \end{align}.

Moreover, by comparing the gauge actions we find that 
\begin{align}\label{MITB} &M_{MN} =\nonumber\\
&S\begin{bmatrix} (2\pi R)^{-2} & -(8\pi^2R^2)^{-1} A_k A^k & -\frac{A_i}{(2\pi R)^2} \\ -(8\pi^2R^2)^{-1} A_k A^k & (2\pi R)^2+(16\pi^2R^2)^{-1} (A_k A^k)^2+A_k A^k  & ((8\pi^2R^2)^{-1} A_k A^k+1)A_i \\ -\frac{A_i}{(2\pi R)^2} & ((8\pi^2R^2)^{-1} A_k A^k+1)A_i & \frac{A_i A_j}{(2\pi R)^2}+\delta_{ij} \end{bmatrix}S, \end{align}
where 
\begin{align}
    S=\frac{1}{\sqrt{2}}\begin{pmatrix}\mathds{1}_{10-d}&\mathds{1}_{10-d}&0\\ \mathds{1}_{10-d}&-\mathds{1}_{10-d}&0\\0&0&\mathds{1}_{r-(10-d)}\end{pmatrix},
\end{align}
and
\begin{align}
    R=2^{-\frac{3}{4}}\mu_8^{-\frac{1}{2}}\left( \int_0^{2\pi} dx^9 w(x^9) \right)^{1/2}  \left( \int_0^{2\pi} dx^9 w(x^9)^{-1}\right)^{-1},
\end{align}
and 
\begin{align} A_i = \frac{1}{2} \left( \int_0^{2\pi} dx^9 w(x^9) \right)^{-1}  \left( \int_0^{x_i^9} dx^9 w(x^9)^{-1} \right).   \end{align}
One can verify that the above results are consistent with those in string theory\footnote{For that, one can first express the moduli of type I' in terms of the Heterotic theory \cite{Polchinski95} and then use eq. 4.7 in \cite{Hohm14} to express the Heterotic moduli in terms of the 9d moduli $M_{MN}$. The parameters $R$ and $A_i$ would be the radius and Wilson lines in the Heterotic picture.}.

According to \eqref{BPSM}, non-BPS limits correspond to limits where the 9d dilaton goes to infinity sufficiently fast. Keeping $M$ fized and sending $\phi$ to infinity corresponds to keeping the relative 8-brane positions fixed while taking $C$ to infinity. If we change the positions of the branes such that $M$ moves on a global geodesic given by \eqref{GGP}, then non-BPS limits correspond to $\Delta\hat\phi \geq(2\sqrt{d-2}/d)|\Delta\gamma|$ which can \textit{always} be accomplished by taking $C \to \infty$ fast enough. We will see that the $C \to \infty$ limit corresponds to a limit of type I' theory where the length of the interval becomes large.

To conclude this section, we will directly show that $C \to \infty$ yields a decompactification limit, verifying that the type I' description of the non-BPS limits is indeed valid. The size of the type I' interval is given by the square root of the metric in the Einstein frame. To convert between the string frame metric $g_{MN}^s$ and the Einstein frame metric $g_{MN}^{\m{E}}$, we have $g^{\m{E}}_{MN} = e^{-\phi_{10}/2} g_{MN}^s$. Thus, 
\begin{align} g_{MN}^{\m{E}} = \tilde{\Omega}^2(x^9) \eta_{MN}, \quad \tilde \Omega = C z^{+1/24} \simeq C^{25/24}. \end{align}
The limit where the interval size goes to infinity is thus $C \to \infty$, and in this limit, we have
\begin{align} m_{\m{KK}} \simeq C^{-25/24}. \end{align}
On the other hand, let us compute the mass scale of the string frame action which we will call the string mass $m_s$. 
\begin{align} m_s = e^{\phi_{10}/4} m_p \simeq C^{-5/24}. \end{align}
We therefore see that $m_{\m{KK}} \ll m_s$ in the limit where the type I' interval becomes large. This is a genuine decompactification limit, as required. 

\section{Gauge symmetry enhancements}\label{sec:gaugesymm}

In the previous section, we showed that the Swampland principles could be used to argue that in certain limits where all BPS states are heavy, the 9d theory must decompactify to a massive IIA background on a compact 1d space with an arrangement of parallel 8-branes. In this section, we will give a bottom-up argument for why the position of the branes uniquely determines the spacetime gauge group. We will show that the gauge groups living on the 8-branes can be determined from Swampland principles.

From string theory, we expect the position of the 8-branes on the interval to encode the information about the gauge theory. This is because the vector multiplets come from open strings ending on the branes, and as two branes approach each other, some massive multiples become massless leading to gauge symmetry enhancement. However, from the bottom-up perspective, the connection between the brane positions and the gauge symmetry is far from clear. Our goal is to provide a bottom-up argument for this connection based on Swampland principles. We provide a bottom-up argument that allows one to read the gauge group directly from the moduli which can be expressed in terms of the brane positions as \eqref{MITB}. For that, we use the recent progress in demonstrating aspects of the string lamppost principle using the Swampland principles. 

In \cite{Hamada:2021bbz,Bedroya21}, it was shown that using the finiteness of black hole entropy and a strong version of the cobordism conjecture, one can classify the possible geometries for the moduli space of the small instantons in minimal supergravities in $d>6$ with 16 supercharges. 

Let us take a moment to explain what small instanton means. For non-Abelian gauge groups, the moduli space of gauge theory instantons can be constructed using the ADHM construction \cite{Atiyah:1978ri}. One of the moduli of the instanton is its size. If we take the size of the instanton to zero. The gauge theory instantons are charged under the ($d-4$)-form gauge symmetry of type IIA and can be instanton. Based on the BPS completeness hypothesis \cite{Kim:2019vuc}, we assume such BPS gauge instantons exist which preserve 8 supercharges. These instantons are known as  non-Abelian small instantons. 

There is a non-Abelian small instanton for any semisimple piece of the gauge algebra. From the field theory perspective, these can be completely disconnected configurations. However, using the Cobordism conjecture \cite{McNamara:2019rup}, in \cite{Bedroya21}, it was argued that one should be able to continuously deform one non-Abelian small instanton to another without breaking the supersymmetry. Therefore, there must exist a BPS co-dimension four defect with a worldvolume theory preserving 8 supercharges, that has a Coulomb branch that parametrizes the non-Abelian small-instantons as well as the BPS configurations that connect them. We refer to this defect as the small instanton.

The non-Abelian instantons correspond to the point of symmetry enhancement on the Coulomb branch of the worldvolume theory of the BPS defect. The global symmetry of the non-Abelian instanton reflects the action of the spacetime gauge transformation on them. So far, the only gauge instantons we considered were non-Abelian, however, it is natural to have instantons even when the spacetime gauge group is broken to its Abelian subgroup. As opposed to the non-Abelian instantons that can be blown up and given a size, the Abelian instantons only exist as point-like objects. For that reason, they are non-perturbative objects whose existence depends on UV information. The field theory construction for them relies on taking some limit in a regularized solution (e.g. non-commutative instantons \cite{Hamanaka:2013vca}). Therefore, the moduli space of the small instanton only has a Higgs branch at specific points of the Coulomb branch where the worldvolume theory has an enhanced non-Abelian global symmetry.

To get a better intuition about the small instantons that connect the gauge instantons, it is helpful to see what they correspond to in string theory. The gauge instantons are typically confined to a stack of branes on which the gauge theory lives. Moreover, different semi-simple pieces of the gauge symmetry correspond to distinct stacks of branes. The BPS configurations that connect the two non-Abelian instantons are D-branes that have four dimensions less than the original branes. The connectedness of this space follows from the fact that an instanton on a stack of brane can be shrunk to zero size and pinched off into a bulk D-brane with four fewer dimensions \cite{Witten:1995gx,Douglas:1995bn}.  Therefore, the small instanton at a generic point of its Coulomb branch is usually a D-brane that can be absorbed into branes with four higher dimensions and create a gauge theory instanton on them. The only conserved quantity that is protected by higher form gauge symmetry is the instanton number plus the corresponding D-brane charge. If they could not transform into each other, we would have conservation of each brane separately, which is not protected by gauge symmetry and leads to a global $(d-4)$-form global symmetry which is forbidden in quantum gravity \cite{Banks:2010zn,McNamara:2019rup}. This implies that in supergravities, the D-brane solutions must be connected to gauge instantons. This implies that co-dimension four D-branes are indeed the small instanton.

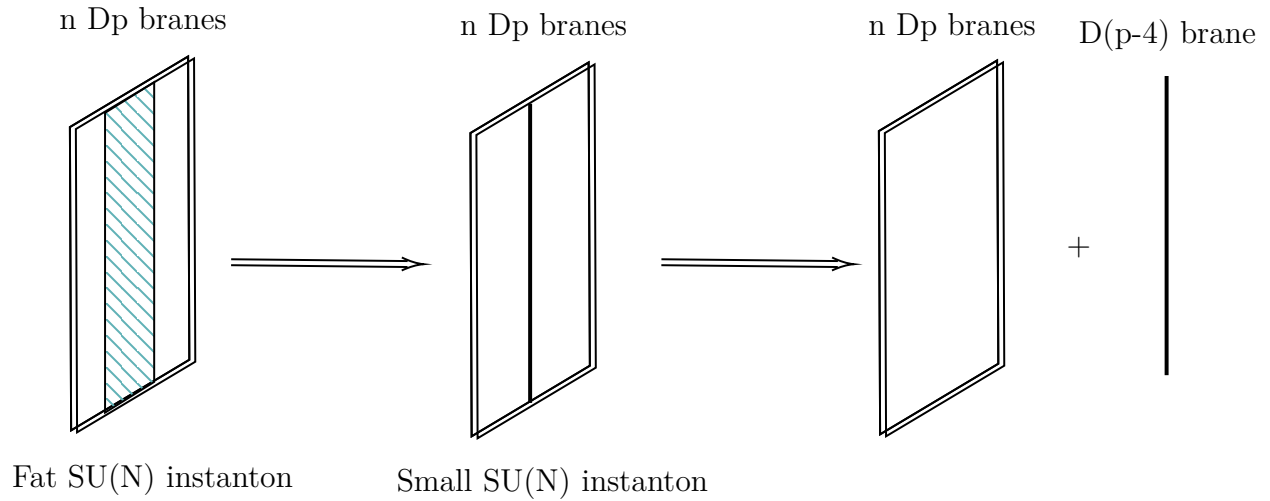
\begin{figure}[H]
    \centering

 
\tikzset{
pattern size/.store in=\mcSize, 
pattern size = 5pt,
pattern thickness/.store in=\mcThickness, 
pattern thickness = 0.3pt,
pattern radius/.store in=\mcRadius, 
pattern radius = 1pt}
\makeatletter
\pgfutil@ifundefined{pgf@pattern@name@_8i1i0kffy}{
\pgfdeclarepatternformonly[\mcThickness,\mcSize]{_8i1i0kffy}
{\pgfqpoint{0pt}{-\mcThickness}}
{\pgfpoint{\mcSize}{\mcSize}}
{\pgfpoint{\mcSize}{\mcSize}}
{
\pgfsetcolor{\tikz@pattern@color}
\pgfsetlinewidth{\mcThickness}
\pgfpathmoveto{\pgfqpoint{0pt}{\mcSize}}
\pgfpathlineto{\pgfpoint{\mcSize+\mcThickness}{-\mcThickness}}
\pgfusepath{stroke}
}}
\makeatother
\tikzset{every picture/.style={line width=0.75pt}} 

\begin{tikzpicture}[x=0.75pt,y=0.75pt,yscale=-1,xscale=1]

\draw   (42.69,74.79) -- (102.3,39.16) -- (102.92,192.23) -- (43.3,227.87) -- cycle ;
\draw   (45.69,75.79) -- (105.3,40.16) -- (105.92,193.23) -- (46.3,228.87) -- cycle ;
\draw   (42.69,74.79) -- (102.3,39.16) -- (102.92,192.23) -- (43.3,227.87) -- cycle ;
\draw  [pattern=_8i1i0kffy,pattern size=6pt,pattern thickness=0.75pt,pattern radius=0pt, pattern color={rgb, 255:red, 74; green, 144; blue, 226}] (60.41,67.61) -- (85.19,52.11) -- (85.19,203.42) -- (60.42,218.91) -- cycle ;
\draw    (124.02,141.5) -- (213.02,142.42)(123.98,144.5) -- (212.98,145.42) ;
\draw [shift={(221,144)}, rotate = 180.59] [color={rgb, 255:red, 0; green, 0; blue, 0 }  ][line width=0.75]    (10.93,-3.29) .. controls (6.95,-1.4) and (3.31,-0.3) .. (0,0) .. controls (3.31,0.3) and (6.95,1.4) .. (10.93,3.29)   ;
\draw   (244.69,77.79) -- (304.3,42.16) -- (304.92,195.23) -- (245.3,230.87) -- cycle ;
\draw   (247.69,78.79) -- (307.3,43.16) -- (307.92,196.23) -- (248.3,231.87) -- cycle ;
\draw   (244.69,77.79) -- (304.3,42.16) -- (304.92,195.23) -- (245.3,230.87) -- cycle ;
\draw [line width=1.5]    (274.8,63.01) -- (274.8,214.01) ;
\draw    (341.02,141.5) -- (430.02,142.42)(340.98,144.5) -- (429.98,145.42) ;
\draw [shift={(438,144)}, rotate = 180.59] [color={rgb, 255:red, 0; green, 0; blue, 0 }  ][line width=0.75]    (10.93,-3.29) .. controls (6.95,-1.4) and (3.31,-0.3) .. (0,0) .. controls (3.31,0.3) and (6.95,1.4) .. (10.93,3.29)   ;
\draw   (450.69,76.79) -- (510.3,41.16) -- (510.92,194.23) -- (451.3,229.87) -- cycle ;
\draw   (453.69,77.79) -- (513.3,42.16) -- (513.92,195.23) -- (454.3,230.87) -- cycle ;
\draw   (450.69,76.79) -- (510.3,41.16) -- (510.92,194.23) -- (451.3,229.87) -- cycle ;
\draw [line width=1.5]    (595.8,49.01) -- (595.8,133) -- (595.8,200.01) ;

\draw (36,13) node [anchor=north west][inner sep=0.75pt]   [align=left] {n Dp branes};
\draw (12,243) node [anchor=north west][inner sep=0.75pt]   [align=left] {Fat SU(N) instanton};
\draw (238,16) node [anchor=north west][inner sep=0.75pt]   [align=left] {n Dp branes};
\draw (206,245) node [anchor=north west][inner sep=0.75pt]   [align=left] {Small SU(N) instanton};
\draw (444,16) node [anchor=north west][inner sep=0.75pt]   [align=left] {n Dp branes};
\draw (544,128.4) node [anchor=north west][inner sep=0.75pt]    {$+$};
\draw (550,19) node [anchor=north west][inner sep=0.75pt]   [align=left] {D(p-4) brane};

\end{tikzpicture}
    \caption{The process of shrinking a gauge instanton on the worldvolume of Dp  branes to zero size and pinching it off into a spacetime $D(p-4) $  brane.}
    \label{POI}
\end{figure}

Now we go back to summarizing the results of \cite{Bedroya21}. Using the cobordism conjecture and finiteness of black hole entropy, the authors showed that the moduli space of small instantons in various dimensions is given as follows\footnote{In all the examples, the resulting manifold matches the internal geometry in a string theory construction, i.e. IIA on interval, F-theory on elliptic K3, and M theory on K3. This is no surprise, given that from string theory we know that the Coulomb branch has a geometric meaning.}:
\begin{itemize}
    \item 9d: An interval with a piece-wise linear profile of $1/g^2$, where $g$ is the coupling constant of the worldvolume theory of the small instanton. The derivative of the coupling constant becomes singular at some points.
    \item 8d: A sphere, with an $SL(2,\mathbb{Z})$-equivariant profile of $\tau$, where $\tau$ is the complexified coupling constant of the worldvolume theory of the small instanton. We can view this configuration as an elliptic K3 surface with some singularities.
    \item 7d: A K3 surface with frozen singularities.
\end{itemize}
Moving in the Coulomb branch of the small instanton corresponds to moving a brane around in the internal geometry. The global symmetries correspond to different singular behaviors in the Coulomb branch of the worldvolume theory of the small instanton. Therefore, one can read off the global symmetries of the small instantons from the geometry of the Coulomb branch. The matching of global symmetries on the small instantons and the gauge symmetries in the spacetime then was used to find a bottom-up classification of ranks and gauge symmetry enhancements in supergravities in $d>6$.

This picture classifies the gauge symmetry enhancements based on the geometry of the moduli space of the small instantons. However, it does not tell us how that geometry depends on the spacetime moduli. This missing piece of information is necessary to understand at which points of the spacetime moduli space the gauge symmetry enhances. 

In this section, we show that our results from the previous section easily fill this gap. In the previous section, we showed that the 9d theory admits a massive type IIA background in specific corners of its moduli space. In these limits, it is easy to see that the compact dimension must represent the Coulomb branch of the 9d small instanton. Above, we used the no-global symmetry conjecture to show that a generic small instanton carries a $(d-5)$-brane charge. Therefore, we are interested in the moduli space of BPS configurations with a BPS 4-brane. Supergravity tells us that placing a 4-brane parallel to the 8-branes and perpendicular to the compact dimension does not break the supersymmetry. Therefore, the position of the 4-brane in the compact dimension is part of the small instanton moduli space in 9d. On the other hand, based on \cite{Bedroya21} we know that the moduli space of small instantons has one real dimension. Therefore, it must correspond to the compact dimension. 

Now we can match the singular point in the moduli space of small instantons  to the position of the 8-branes. Then we can use the results of \cite{Bedroya21} to read off the spacetime gauge symmetry. Therefore, we can find the gauge symmetry in the spacetime by finding the position of the 8-branes in type I' supergravity. For example, in a generic point of the moduli space where the gauge group is $U(1)^r$, there must be $r+1$ 8-branes along the interval whose positions are constrained to make sure that dilaton goes to infinity at the two endpoints of the interval. As 8-branes approach each other and coincide, the gauge symmetry enhances. We can see the enhanced gauge symmetry, by reading off the global symmetry of the small instanton that has the corresponding singular structure in \cite{Bedroya21}. For example, in the maximal rank case of $r=17$, if we move 8 8-branes to each endpoint and leave two at the center, the gauge algebra will be $e_8+e_8+su(2)$.

Note that if the gauge group has an $E_n$ piece,  the worldvolume theory of the small instanton at the endpoint of its Coulomb branch will have $E_n$ global symmetry. From the classification of 5d SCFTs we know that the coupling must diverge at that point. Therefore, following the above mapping between the internal geometry and the moduli of the small instanton, we conclude that in such cases, the dilaton must diverge at the endpoints of the interval.

This approach also provides a Swampland argument for the fact that the world volume theory of a stack of 8-branes has a $U(n)$ gauge group. Moreover, we can read off the gauge group living on the end of the universe wall, and the top-form charge that brane must have to cancel the flux of 8-branes.  By fixing the number of the 8-branes, this argument even fixes the top-form charge of the end of the universe wall and  we can even read off the gauge group living on the stack of 8-branes on top of an end of the universe branes, which is non-trivial.

Our argument maps the gauge symmetry to the position of the branes, and the brane positions can be expressed in terms of the 9d moduli. Therefore, we know the location of the gauge symmetry enhancements in terms of the 9d moduli. On the other hand, the dependence of the gauge symmetry enhancement on the 9d moduli is controlled by the charge lattice. The dependence of the charge lattice on the moduli is controlled by supersymmetry\footnote{See Appendix D in \cite{Bedroya:2023}} and whenever the sublattice of zero charges under the graviphoton has an ADE sublattice, we have a gauge symmetry enhancement. Therefore, our argument uniquely fixes the charge lattice. Since the charge lattice is even self-dual in string theory constructions in 9d, we find that the choice is unique. In \cite{Bedroya21}, all gauge symmetry enhancements of the lattice were classified and shown to belong in the same moduli space. This strongly suggests that the charge lattice which gives rise to those symmetry enhancements is unique. Our work provides a concrete explanation for the uniqueness of the charge lattice but also gives a bottom-up explanation for the strong coupling behavior of the infinite distance limit.

\section{Extension to lower dimensions}\label{sec6}

In this section, we comment on the applicability of our methods to lower dimensions. One might expect that as long as we know that the theory decompactifies, we can still identify the moduli space of small instantons with the internal geometry. In this section, we show that the connection between two spaces is generally non-trivial. For example, consider 8d theories with 16 supercharges. Two possible decompactification limits, which are also non-BPS limits, are the 32 supercharges supergravity on an interval ($8+1-d $ analogue of type I') and type IIB on $\mathds{P}^1$ with a varying axion-dilaton\footnote{Such backgrounds are only possible due to the $SL(2,\mathbb{Z})$ S-duality of the type IIB which allows for monodromies of the axio-dilaton. The S-duality is not trivial from low energy theory. However, \cite{Bedroya:2023} provides a bottom-up argument for this duality using Swampland principles.}.

As we reviewed in the previous section, the moduli space of the small instanton in 8d is $\mathds{P}^1$. From string theory, we expect this $\mathds{P}^1$ to be the internal geometry in the type IIB background. In fact, if the theory decompactifies to a type IIB background, we can follow an argument very similar to that of the 9d case to match the two $\mathds{P}^1$'s and find a bottom-up derivation for the gauge symmetry on the 7-branes. The argument is based on the fact that a generic small instanton in 8d must carry a 3-brane charge. Therefore, the moduli space must match that of a 3-brane moving in the $\mathds{P}^1$ base in type IIB background.

Now let us look at the other possibility, which is supergravity on an interval. As we will explain using a top-down picture in string theory, the connection between this interval and the moduli space of the small instantons is non-trivial and obscured to low-energy physics. In string theory, this background arises as type IIA theory on $S^1$, which is further compactified on an interval. However, this picture obscures the small instanton moduli space. The lower dimensional small instantons are the  D3-branes which correspond to wrapped D4-branes. From the local T-dual IIB picture, we know that such D3-branes have a compact modulus corresponding to the location of the brane in the circle. There are also D7-branes along the interval that come from the wrapping of the D8 branes in the type IIA picture. Similar to the D3-branes, the D7-branes also carry a compact modulus. Therefore, we roughly expect a 2d Coulomb branch which comes from having a circle on top of every point in the interval, with some singularities near the location of 7-branes. This matches our expectation of having a $\mathds{P}^1$ moduli space. However, to reconstruct the moduli space of small instantons, the interaction between the D3-brane and the D7-branes becomes important. From the study of the worldvolume theory of the instanton, we know that the singularities are pointlike \cite{Bedroya21}. The singularity structure arises from the interplay between the compact modulus of D3- and D7-branes. If we move the D3-brane around, both in the interval and in its $S^1$ Coulomb branch, its theory becomes singular when it meets a D7-brane with a matching compact modulus. One can see this much more clearly in the type IIB picture where the compact moduli correspond to the location of D3 and D7 in $S^1$ fiber of $\mathds{P}^1$. The D3 brane only hits the D7 brane if their locations on the $S^1$ are synced up. However, seeing this from the lower dimensional theory is difficult since it will require the low energy theory of a D3 brane probing the D7 brane. For a detailed description of how the type I' construction can be understood from the type IIB picture see \cite{Cachazo:2000ey}. Therefore, even in 8 dimensions, if we decompactify an interval, it is challenging to reconstruct the moduli space of the small instanton because only a part of the moduli space has found a geometric meaning.

\begin{figure}[H]
    \centering

\tikzset{every picture/.style={line width=0.75pt}} 

\begin{tikzpicture}[x=0.5pt,y=0.5pt,yscale=-1,xscale=1]

\draw  [fill={rgb, 255:red, 0; green, 0; blue, 0 }  ,fill opacity=0.27 ] (148.69,124.35) -- (208.31,95.8) -- (208.92,218.44) -- (149.3,247) -- cycle ;
\draw [color={rgb, 255:red, 74; green, 144; blue, 226 }  ,draw opacity=1 ][line width=1.5]    (326.8,108.5) -- (326.8,175.79) -- (326.8,229.49) ;
\draw    (113,173.39) -- (149.5,173.39) ;
\draw [shift={(113,173.39)}, rotate = 180] [color={rgb, 255:red, 0; green, 0; blue, 0 }  ][line width=0.75]    (0,5.59) -- (0,-5.59)   ;
\draw    (185,173.39) -- (523.5,172.59) ;
\draw [shift={(523.5,172.59)}, rotate = 179.86] [color={rgb, 255:red, 0; green, 0; blue, 0 }  ][line width=0.75]    (0,5.59) -- (0,-5.59)   ;
\draw  [dash pattern={on 4.5pt off 4.5pt}]  (176.5,67.44) -- (176.5,100) ;
\draw   (153,37) .. controls (153,23.19) and (164.19,12) .. (178,12) .. controls (191.81,12) and (203,23.19) .. (203,37) .. controls (203,50.81) and (191.81,62) .. (178,62) .. controls (164.19,62) and (153,50.81) .. (153,37) -- cycle ;
\draw  [fill={rgb, 255:red, 0; green, 0; blue, 0 }  ,fill opacity=1 ] (155,50.75) .. controls (155,49.23) and (156.23,48) .. (157.75,48) .. controls (159.27,48) and (160.5,49.23) .. (160.5,50.75) .. controls (160.5,52.27) and (159.27,53.5) .. (157.75,53.5) .. controls (156.23,53.5) and (155,52.27) .. (155,50.75) -- cycle ;
\draw  [dash pattern={on 4.5pt off 4.5pt}]  (326.5,66) -- (326.5,98.56) ;
\draw  [color={rgb, 255:red, 74; green, 144; blue, 226 }  ,draw opacity=1 ] (303,34) .. controls (303,20.19) and (314.19,9) .. (328,9) .. controls (341.81,9) and (353,20.19) .. (353,34) .. controls (353,47.81) and (341.81,59) .. (328,59) .. controls (314.19,59) and (303,47.81) .. (303,34) -- cycle ;
\draw  [fill={rgb, 255:red, 74; green, 144; blue, 226 }  ,fill opacity=1 ] (322.5,59) .. controls (322.5,57.48) and (323.73,56.25) .. (325.25,56.25) .. controls (326.77,56.25) and (328,57.48) .. (328,59) .. controls (328,60.52) and (326.77,61.75) .. (325.25,61.75) .. controls (323.73,61.75) and (322.5,60.52) .. (322.5,59) -- cycle ;

\draw (174,255.4) node [anchor=north west][inner sep=0.75pt]    {$D7$};
\draw (317,258.4) node [anchor=north west][inner sep=0.75pt]    {$D3$};

\end{tikzpicture}
    \caption{The 9d maximal supergravity on an interval with 
BPS domain walls along the interval. In string theory, the domain walls correspond to 7 branes, and the 8d small instanton is the D3 brane. Each brane has a compact modulus corresponding to its location in the circle in the type IIB frame. Therefore, the moduli space of the small instanton is a circular fibration over the interval, which is a $\mathds{P}^1$. The D3 brane probes the singularities on the $\mathds{P}^1$ when both its position and its compact modulus approach those of the D7 branes.}
    \label{fig:my_label}
\end{figure}
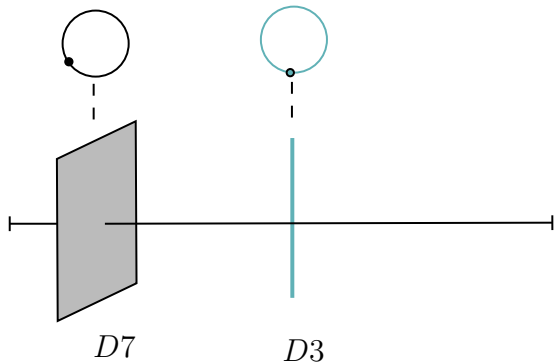

Note that this does not happen in the 9d case because the real dimension of the moduli space of the small instantons is one. Therefore, the full moduli space finds a geometric meaning when we decompacity to a type I' background.

\section{Conclusion}

We used the strong version of the sharpened distance conjecture \cite{Etheredge:2022opl} and the finiteness of black hole entropy to show that any non-BPS limit of a 9d supergravity must decompactify to IIA supergravity on a one-dimensional manifold. Then, we used the classification of 5d SCFTs and no-global symmetry conjecture \cite{Banks:2010zn} to show that the internal geometry must be an interval with BPS 8-branes placed along the interval. We also showed that the internal geometry must match the moduli space of the 9d small instantons. This allowed us to read the gauge group from the positions of the branes which are directly expressed in terms of the 9d moduli using supersymmetry alone. Therefore, we find a bottom-up argument for the low-energy worldvolume theory of an arbitrary stack of non-perturbative 8-branes. 

Our argument demonstrates the power of the sharpened distance conjecture \cite{Etheredge:2022opl}. Usually, the BPS limits are preferred in bottom-up studies due to having more control over the light states. However, if one assumes the sharpened distance conjecture, we showed that the non-BPS limits could be equally powerful. If we know that the theory must decompactify, the fact that light particles are non-BPS significantly reduces the options for the internal geometry. 

In \cite{Bedroya:2023}, the ideas and results of this work are extended to provide a bottom-up argument for all string dualities between 10d and 11d supergravity backgrounds. It would be interesting to see if the blend of ideas we presented here could be extended to lower dimensions and motivate the string lamppost principle even further. However, as we discussed in section \ref{sec6}, this pursuit gets quickly complicated. We were informed of an upcoming work which also studies the type I' limits in light of the Swampland distance conjectures \cite{irene}.

\section*{Acknowledgement}

We thank  Yuta Hamada, Jacob McNamara, and Cumrun Vafa for insightful discussions and helpful comments. The research of A. B. and H. C. T. is supported by a grant from the Simons Foundation (602883,CV) and by the NSF grant PHY-2013858. 

\bibliographystyle{unsrt}
\bibliography{bib}

\end{document}